\documentclass[reqno,11pt]{amsart}
\usepackage[left=1in,right=1in,top=1in,bottom=1in]{geometry}

\usepackage{amssymb,amsfonts,mathrsfs,stmaryrd}
\usepackage[final]{microtype}
\usepackage{times}

\usepackage{dsfont}
\usepackage{booktabs}
\usepackage[utf8]{inputenc} 
\usepackage[T1]{fontenc}
\usepackage{url}

\usepackage{parskip}

\theoremstyle{plain}
\newtheorem{theorem}{Theorem}
\newtheorem{proposition}[theorem]{Proposition}

\newtheorem{definition}[theorem]{Definition}

\theoremstyle{remark}
\newtheorem{remark}[theorem]{Remark}

\newcommand{\eqdef}{\triangleq}
\renewcommand{\leq}{\leqslant}

\newcommand{\set}[1]{\mathcal{#1}}
\newcommand{\rg}[2]{\llbracket #1,#2  \rrbracket}

\newcommand{\sym}{\mathfrak{S}}

\newcommand{\F}{\mathbb{F}}
\newcommand{\Z}{\mathbb{Z}}

\newcommand{\fp}{\F_{p}}
\newcommand{\fpm}{\F_{p^m}}
\newcommand{\fq}{\F_{q}}

\newcommand{\scp}[2]{\left \langle #1 , #2  \right\rangle} 
\newcommand{\hscp}[3]{\left \langle #2 , #3  \right\rangle_{#1}} 

\newcommand{\word}[1]{\boldsymbol{#1}}
\newcommand{\av}{\word{a}}
\newcommand{\ev}{\word{e}}
\newcommand{\uv}{\word{u}}
\newcommand{\vv}{\word{v}}
\newcommand{\xv}{\word{x}}
\newcommand{\yv}{\word{y}}
\newcommand{\zv}{\word{z}}
\newcommand{\zz}{\word{0}}

\newcommand{\dual}[1]{{#1}^\bot}
\newcommand{\hdual}[2]{{#2}^{\bot_{#1}}}
\newcommand{\dualp}[1]{\left(#1\right)^\bot}
\newcommand{\hull}[1]{{\mathcal{H}\left(#1\right)}}
\newcommand{\sh}[2]{{\mathcal{S}}_{#1}\left(#2\right)}

\newcommand{\AC}{A}
\newcommand{\BC}{B}

\DeclareMathOperator{\rank}{\mathsf{rank}}
\DeclareMathOperator{\im}{\mathsf{Im}}

\newcommand{\mat}[1]{\boldsymbol{#1}}
\newcommand{\Am}{\mat{A}}
\newcommand{\Sm}{\mat{S}}
\newcommand{\Gm}{\mat{G}}
\newcommand{\Hm}{\mat{H}}
\renewcommand{\Im}{\mat{I}}
\newcommand{\Xm}{\mat{X}}
\newcommand{\Zm}{\mat{Z}}

\newcommand{\Sim}{\mat{\Sigma}}

\newcommand{\adj}[1]{\Am\left({#1} \right)}
\newcommand{\graph}[1]{\mathcal{#1}}
\newcommand{\GI}{{\mathsf{GI}}}

\newenvironment{IEEEproof}{\paragraph{\textbf{Proof.}}}{\hfill$\blacksquare$}

\title{\textbf{Permutation Code  Equivalence is Not Harder Than  Graph Isomorphism When Hulls Are Trivial}}

\author{Magali Bardet
\and 
Ayoub Otmani
\and
Mohamed  Saeed-Taha}
\thanks{M. Bardet, A. Otmani and M. Saeed-Taha are with Normandie Univ, France; UR,  LITIS, F-76821 Mont-Saint-Aignan, France. M. Saeed-Taha is also with University of Khartoum, Sudan}

\date{}

\begin{document}

\maketitle
\begin{abstract}
The paper deals with the problem of deciding if
 two  finite-dimensional linear subspaces over  an arbitrary  field are identical  up to a permutation of the coordinates.  This problem is referred to as the \emph{permutation code equivalence}.
We show that given access to a subroutine that decides if two weighted undirected graphs are isomorphic, one may deterministically decide the permutation code equivalence provided that the underlying vector spaces intersect trivially with their orthogonal complement with respect to an arbitrary inner product.  Such a class of vector spaces is usually  called linear codes with trivial hulls. The reduction is efficient because it essentially boils down to computing the inverse of a square matrix of order the length of the involved codes. 
Experimental results obtained with randomly drawn binary codes having trivial hulls show that permutation code equivalence can be decided in a few minutes for lengths up to $50,000$.
\end{abstract}

\section{Introduction} \label{sec:intro}

This paper deals with the problem of deciding whether two  finite dimensional linear subspaces of $\F^n$ where $\F$ is a field (not necessarily finite) are identical  up to a permutation of the coordinates.  This problem which is referred to as the \emph{permutation code equivalence}  is a special case of the \emph{isomorphism problem} that asks, 
 for a given \emph{metric}, whether  there exists a linear \emph{isometry}
between two linear subspaces.  These problems are usually encountered in coding theory where 
the ambient space $\F^n$ is  endowed with  the Hamming distance.
The  isomorphism problem is  of great importance in the classification of codes   because isomorphic codes display identical characteristics such as length, dimension, minimum distance, weight distribution and 
correction capabilities. The permutation code equivalence problem was  used in cryptography by McEliece in \cite{M78} as a tool for improving the security of public key encryption schemes. Although the security of the McEliece scheme is not directly related to the permutation equivalence problem, there exist some encryption schemes that can be fully 
cryptanalyzed if an attacker efficiently solves the  equivalence problem (see for instance
\cite{MS07,BCDOT16}).
Consequently a better assessment of the computational difficulty of this problem  will permit to 
more accurately evaluate the  security of code-based cryptographic primitives.

Its computational difficulty was studied  in \cite{PR97} where  it was  shown that if the permutation equivalence problem  over the binary field is NP-complete, then the polynomial hierarchy collapses. Hence this problem is unlikely to be NP-complete. 
They also provided   a polynomial-time reduction from the graph isomorphism  to the 
permutation equivalence problem.  However, in light of the recent breakthrough  showing that graph isomorphism
might not be a hard problem \cite{Babai15},  this reduction becomes less  interesting.
A worst-case upper-bound was given in \cite{BCGQ11} showing that  the permutation equivalence problem can be solved with $(2+o(1))^n$ operations. 

Another very important question connected to the isomorphism problem consists  in characterizing the group of isometries  that leaves globally invariant a given  linear  space. 
This group is called the \emph{automorphism group}.
The set of isomorphisms between two equivalent linear subspaces is then  a coset of the automorphism group. In particular 
the number of solutions (when they exist) to the isomorphism problem is exactly  the size of the 
automorphism group.
Since  permutations are always isometries for \emph{any} distance over $\F^n$
the set of permutations that  globally stabilize  a linear space is called the \emph{permutation group}.
Leon  \cite{L82}  introduced a method  for fully determining the permutation group of a linear code over 
a finite field $\fq$. 
The algorithm requires the computation of all codewords of a given weight. The time complexity is therefore
 $n q^{O(k)}$  for $k$-dimensional vector subspaces assuming that the time complexity of  field operations 
 are constant. The techniques developed in Leon's algorithm can then be used to solve the permutation equivalence problem for linear codes of small dimension over small  fields. This algorithm is for instance implemented in Magma software \cite{BCP97}.
Later  Sendrier    \cite{S00} considered  the  \emph{hull} of codes which is the intersection of a vector subspace with its orthogonal complement with respect to the Euclidean inner product. 

The  underlying motivation is that 
 (permutation) equivalent linear subspaces have permutation equivalent hulls. 
 Although the permutation problem between hulls may have more solutions,  and moreover inequivalent codes may have equivalent hulls,
  the practical advantage  is   that hulls tend to have very small dimensions \cite{s97a}. Actually
  hulls are very likely to be reduced to $\{ \zz \}$ which paradoxically does not help in solving the  problem.
 Consequently Sendrier's approach  is interesting when the hull is very small but not trivial, and
its permutation group  is ideally trivial. 
The technique developed in \cite{S00} first punctures at each position the code to be tested, and
then computes for each punctured code the weight enumerator of its hull.
All these weight enumerators provide  a compact  ``signature'' 
with the property that permutation 
equivalent codes have equivalent signatures.   The permutation 
equivalence problem is therefore reduced to testing whether two signatures are 
equivalent. The time complexity of \cite{S00} is  at least exponential in the dimension of the hull.
Furthermore, 
 in order to take into consideration the cases where there may exist many solutions, 
 not to mention  the fact that two inequivalent codes may have equivalent signatures,
several successive puncturing operations have to be applied, which complicates the analysis of the algorithm. 
However it is claimed (see for example \cite{SS13}) that its empirical time complexity is $O\left(n^3 + 2^h n^2 \log n\right)$ where $h$ is the dimension of hull, which means 
that  the algorithm becomes polynomial  when restricted to codes with hulls of bounded dimension.

Recently a new algebraic approach was introduced  in \cite{S17}  in order to solve the permutation code equivalence.  
It builds a quadratic polynomial system whose binary solutions are permutations.  It was then shown 
that Gr\"obner basis techniques fully solve the code equivalence but the cost becomes rapidly prohibitive as the length of the codes increases. Several improvements are then proposed enabling  to efficiently solve the permutation equivalence when the codes have small hulls or are defined over large fields.

These previous works tend to prove that the hardness of the permutation code equivalence is not well understood. From a practical point of view, there is a prevalent belief that this  problem is easy  (see for example \cite{S02}) when dealing with codes having very small hulls, as it is the case for random linear codes. 
On the other hand, no rigorous theoretical bound supporting these concrete observations  
is known, especially there is no proof showing that permutation code equivalence is polynomial
 in time for codes with trivial hulls. Furthermore several works are dedicated to design practical algorithms \cite{L82,S00,B07,F09b} but  none seems to be able to decide efficiently if two  codes  of rate say one-half and of length  greater than $200$ are equivalent or not. The situation is getting worse when the fields are large, not to mention fields of characteristic $0$,  because  all these methods require small finite fields.

\subsection*{Our contributions}

We show that given access to a subroutine that decides if two weighted undirected graphs are isomorphic, we may deterministically decide the permutation code equivalence over an arbitrary field $\F$ assuming that the hulls of the involved vector spaces are trivial.
Our reduction given in Theorem~\ref{theo:red} is very efficient since it essentially boils down to computing the inverse of a square matrix of order $n$ where $n$ is the length of the codes considered. 
We exploit the fact that the direct sum of a vector subspace $U \subset \F^n$  and its orthogonal $\dual{U} $ is equal to 
the full vector space $\F^n = U \oplus \dual{U}$ when $U \cap \dual{U} = \{\zz \}$.
It is then possible to decompose each vector $\ev_i$ of  the canonical basis $\big\{ \ev_i :  i \in \rg{1}{n} \big\}$ as 
$\ev_i = \sigma_U (\ev_i) + \sigma_{\dual{U}} (\ev_i)$ where $\sigma_U (\ev_i)$ (\textit{resp.} $\sigma_{\dual{U}} (\ev_i)$) is the projection of $\ev_i$ on $U$ (\textit{resp.} $\dual{U}$).
We then define the square matrix $\Sim_U$ whose rows are formed by the projections $\sigma_U (\ev_i)$.
In the same way, we define $\Sim_{\dual{U}}$ which enables to write $\Sim_{U} + \Sim_{\dual{U}} = \Im_n$. We prove that 
$\Sim_{U}$ and  $\Sim_{\dual{U}}$ are symmetric and for any permutation matrix  $\Xm$ it holds that
 $\Xm^T \left (\Sim_{U} + \Sim_{\dual{U}} \right) \Xm =  \Xm^T\Sim_{U} \Xm + \Xm^T\Sim_{\dual{U}}\Xm$. By the uniqueness of the decomposition, it entails that  $\Sim_{U \Xm} = \Xm^T\Sim_{U} \Xm$ and
 $\Sim_{\dual{U} \Xm} = \Xm^T\Sim_{\dual{U}}\Xm$. These relations express the property 
 that graphs represented by $\Sim_{U \Xm}$ and  $\Sim_{U}$ viewed as adjacency matrices are isomorphic.

The advantage of this reduction is twofold. From a practical point of view, the permutation equivalence would benefit from the (numerous) efficient existing tools that solve the problem of isomorphism between graphs.  
Our experimentations with binary codes of length up to $20,000$ show that we can decide
if two codes are equivalent in less than two seconds. Our computations were performed on 
 Intel Core i7, 2.5 GHz, 16GB ram  with Magma software 
\cite{BCP97} which relies on version  2.6r7 of the \texttt{nauty} and \texttt{Traces} packages \cite{MP14}.
The other advantage is that,  in light of the recent result obtained in \cite{Babai15}, our reduction validates in a sense  the  belief  that codes with trivial hulls form a class of easy instances for the permutation code equivalence problem.

Finally, we propose two solutions to treat the  cases of  codes  having a non-trivial intersection with their orthogonal complement. The first one applies more particularly to codes defined over an extension field $\fpm$ where $m > 1$ and $p$ is a prime. We equip the space $\fpm^n$ with an Hermitian inner product
 by means of an automorphism of $\fpm$. This enables us to adopt a different perspective because 
a code may not have  a non-trivial hull for one inner product but has a trivial hull with respect to another one. 
With little effort, we are able to generalize our reduction  to any Hermitian inner  product. This generalization  permits  to encompass  all but an exponentially small fraction of linear codes.
The remaining cases are then essentially codes defined over a prime field $\fp$. By shortening sufficiently enough it is possible to obtain codes with trivial hulls.  Given access to an oracle that decides the graph isomorphism, we then propose an algorithm whose time complexity is  $O\Big(   h n^{\omega+h+1} \GI(n)     \Big)$ where $\omega$ is the exponent of matrix multiplication ($2 \leq \omega < 3$) and
$\GI(n)$  is  the time complexity for deciding if two weighted graphs with $n$ vertices and weights in $\F$ are isomorphic.

The rest of the paper is organized as follows.
Section~\ref{sec:prelim} introduces notation and classical notions on algebra and graph theory. 
It also introduces  the permutation code equivalence problem. Section~\ref{sec:red} is dedicated to the presentation of 
a deterministic reduction from the permutation code equivalence to graph isomorphism.
Section~\ref{sec:non)trivial-hull} is concerned with methods that deal with codes having non-trivial hulls.

\section{Preliminaries} \label{sec:prelim}

\subsection{Notation} 

Throughout the paper $\F$ is an arbitrary  field.
For any $m$ and $n$ in $\Z$ the notation $\rg{m}{n}$ stands  for the
set of integers $i$ such that $m \leq i \leq n$.
The (Euclidean) \emph{inner product} between $\xv$ and $\yv$ from $\F^n$ is 
$\scp{\xv}{\yv} \eqdef \sum_{i=1}^n x_i y_i$. 
$\sym_n$ is the  \emph{Symmetric} group on $\rg{1}{n}$. 
The action of $\sym_n$ over $\F^n$
is defined for any $\pi$ in $\sym_n$ and $\uv$ in $ \F^n$ by
$
\uv^\pi \eqdef   \left( u_{\pi(1)},\dots{},u_{\pi(n)} \right)$.
We extend this notation  to any subset $U \subseteq \F^n$, namely 
$U^\pi \eqdef \left \{  \uv^\pi : \uv \in U\right \}$.
 Finally, we will rely on the fact that the computation of the inverse of an $n \times n$ matrix over a field $\F$ 
can be performed in $O(n^\omega)$ operations  assuming that the operations in $\F$ are constant where
$\omega$ is the exponent of matrix multiplication ($2 \leq \omega < 3$). 
The best theoretical bound is currently $\omega < 2.373$.

\subsection{Permutation code equivalence} 

A \emph{linear code $U$ of length} $n$ and \emph{dimension} $k$ over a field $\F$ is a $k$-dimensional vector subspace of $\F^n$.  
Any $k \times n$ matrix whose rows form a basis is 
called a \emph{generator matrix}. 
The \emph{orthogonal complement} of a code $U \subset \F^n$
 is the linear space $\dual{U}$ containing  all
vectors $\zv$ from $\F^n$ such that for all $\uv \in U$, we have $\scp{\uv}{\zv} = 0$.
We always have $\dim \dual{U} = n - \dim U$, and
any generator matrix of $\dual{U}$ is called a \emph{parity check} matrix of $U$.
Throughout the paper  we use the convention that $\Gm_U$ and $\Hm_U$ 
represent  respectively a generator matrix and a parity check matrix of a linear code $U$.
The  \emph{Hull}   of  a code $U \subset \F^n$ is  $ U \cap \dual{U}$. A parity-check matrix of $ U \cap \dual{U}$
is then \[
\Hm_{U \cap \dual{U}} \eqdef \begin{bmatrix} \Gm_U \\ \Hm_U\end{bmatrix}.
\]
When $ U \cap \dual{U} = \{ \zz \}$ we say that 
$U$ has a trivial hull or is a \emph{linear code with a complementary dual}. 
It means  that both $\Gm_U \Gm_U^T$ and $\Hm_U \Hm_U^T$ are invertible. 
It is also equivalent to say that $\F^n = U \oplus \dual{U}$, that is to say 
$\Hm_{U \cap \dual{U}}$ is invertible with
\begin{equation} \label{eq:Hm_1}
\Hm_{U \cap \dual{U}}^{-1} = 
	\begin{bmatrix}
 		\Gm_U^T \Big( \Gm_U \Gm_U^T \Big)^{-1} &  \Hm_U^T \Big( \Hm_U \Hm_U^T \Big)^{-1} 
 	\end{bmatrix}.
\end{equation}

Two linear codes  $\AC$ and $\BC$ of length $n$ are  \emph{permutation equivalent} if there
exists $\pi$ in $\sym_n$ such that $\BC = \AC^\pi$. We shall use the notation
$\BC \sim_\pi \AC$.
This is also equivalently denoted by $\BC = \AC \Xm \eqdef \left \{ \av \Xm : \av \in \AC \right \}$ where $\Xm$ is a
 matrix representing the permutation $\pi$.

\begin{definition}
Given two linear codes $\AC$ and $\BC$ of length $n$, the  permutation code equivalence  
problem asks if there exists a permutation $\pi$ in $\sym_n$ such that $\BC = \AC^\pi$.
\end{definition}

\begin{remark}
Two linear codes $\AC$ and $\BC$ are permutation equivalent if and
  only if  $\Gm_{\BC} = \Sm \Gm_{\AC}\Xm$
  where $\Sm$ is an invertible matrix and $\Xm$ is a permutation matrix. Furthermore
it is not difficult to see that $\BC \sim_{\pi}  \AC$  if and only if
 $\dual{\BC} \sim_{\pi}  \dual{\AC}$. 
\end{remark}

\subsection{Graph theory}

A \emph{weighted} (or \emph{edge-labeled}) graph $\graph{G} = (V, E)$ is composed of a finite set of \emph{vertices} 
$V$,  a set of \emph{edges} $E \subset V \times V$, a set of weights $W$, and a function which assignes a weight $w(i,j)$ to each edge $(i,j) \in E$. 
 We will always assume that $V = \rg{1}{n}$.  The graph $\graph{G}$ is \emph{undirected} if 
for each $(i,j)$ in $E$, we also have $(j,i)$ in  $E$ and $w(i,j) = w(j,i)$. 
The \emph{adjacency} matrix $\adj{\graph{G}} = \begin{bmatrix} a_{i,j} \end{bmatrix}$ 
of a graph $\graph{G}$ is an $n \times n$ matrix such that $a_{i,j} = w(i,j)$ when $(i,j) \in E$ and 
$a_{i,j} = 0$ otherwise. In particular $\adj{\graph{G}}$ is \emph{symmetric} when $\graph{G}$ is undirected.
The \emph{graph isomorphism} (GI) problem is given two graphs $\graph{G}_1 = (V,E_1)$ and $\graph{G}_2 = (V,E_2)$ with $n$ vertices, determine whether $\graph{G}_2$ can be obtained from $\graph{G}_1$ by permutation of its vertices while keeping both adjacent vertices \emph{and} weight edges. It is equivalent to determine if there exists a permutation $\pi$ in $\sym_n$ such that  
$(\pi(i),\pi(j))$ in $E_2$ if and only if $(i,j)$ in $E_1$ and $w\big( \pi(i), \pi(j) \big) = w(i,j)$. We shall write in that case 
$\graph{G}_2 \sim_\pi \graph{G}_1$. 
This also means that there exists an $n \times n$ permutation matrix $\Xm$ such that 
\begin{equation} \label{eqGI}
\adj{\graph{G}_2} = \Xm^T \adj{\graph{G}_1} \Xm.
\end{equation}

\section{A Reduction To Weighted Graph Isomorphism} \label{sec:red}

Let us assume that $U \subset \F^n$ is  a code with a trivial hull, and consequently we have $\F^n = U \oplus \dual{U}$. 
Let $\sigma_U : \F^n \rightarrow U$ and $\sigma_{\dual{U}} : \F^n \rightarrow \dual{U}$ be the 
projections associated to $U$ and $\dual{U}$ respectively. 
Any vector $\vv$ in $\F^n$ can be uniquely written as 
$\vv = \sigma_U(\vv) + \sigma_{\dual{U}}(\vv)$. 
The linear maps $\sigma_U$  and $\sigma_{\dual{U}}$
satisfy the following relations: $\sigma_U + \sigma_{\dual{U}} = id$, $\sigma_U^2 = \sigma_U$, $\sigma_{\dual{U}}^2 = \sigma_{\dual{U}}$ and
$\sigma_{\dual{U}} \circ \sigma_U= \sigma_U \circ \sigma_{\dual{U}} = \zz$. We also have 
that $\im(\sigma_U) = U$ and $\im(\sigma_{\dual{U}}) = \dual{U}$.
 In order to characterize matrices  defining $\sigma_U$  and $\sigma_{\dual{U}}$, we introduce
 the following $n \times n$ matrices 
\begin{equation} \label{eq:SigU}
\left \{
\begin{array}{lcl}
\Sim_U &\eqdef& \Gm_U^T \Big( \Gm_U \Gm_U^T \Big)^{-1} \Gm_U, \\
\Sim_{\dual{U}} &\eqdef&   \Hm_U^T \Big( \Hm_U \Hm_U^T \Big)^{-1} \Hm_U.
\end{array}
\right.
\end{equation}

\begin{proposition} \label{prop:hulldecompose}
Let $U$ be a linear code with trivial hull then  it holds that
\begin{equation}
\forall \vv \in \F^n, \;\;\; \sigma_{U}(\vv) = \vv \, \Sim_U
~~~~\text{ and }~~~~
 \sigma_{\dual{U}}(\vv) = \vv \,  \Sim_{\dual{U}}.
\end{equation}
\end{proposition}

\begin{IEEEproof}
Let us assume that $\dim U = k$, and let $\vv$ be an arbitrary element from $\F^n$.
Finding $\sigma_U(\vv)$ and $\sigma_{\dual{U}}(\vv)$ is equivalent to 
finding $\xv_{U}$ from $\F^k$ and  $\xv_{\dual{U}}$  from $\F^{n-k}$ 
such that $\sigma_U(\vv) =   \xv_{U} \Gm_U$  and $\sigma_{\dual{U}}(\vv) =  \xv_{\dual{U}} \Hm_U$, which means that 
$\vv = \left ( \xv_U, \xv_{\dual{U}} \right) \Hm_{U \cap \dual{U}}$. By assumption $\Hm_{U \cap \dual{U}}$ 
 is invertible and by using   \eqref{eq:Hm_1}  we obtain then  
   $\xv_{U} =  \vv \; \Gm_U^T \Big( \Gm_U \Gm_U^T \Big)^{-1}$ and
   $\xv_{\dual{U}} = \vv \;  \Hm_U^T \Big( \Hm_U \Hm_U^T \Big)^{-1}$.
\end{IEEEproof}

  Note that $\Sim_U$ only depends on the code $U$ and not on a
  particular generator matrix  of $U$ since for any invertible matrix $\Sm$ it holds that
  \begin{equation}
  \left( \Sm\Gm_U \right)^T \Big( \Sm \Gm_U \left( \Sm \Gm_U \right)^T \Big)^{-1} \left( \Sm \Gm_U \right)
  =
  \Sim_U
  \end{equation}
  Moreover $\Sim_U$ (and $\Sim_{\dual{U}}$) satisfies  the relations  $\im\left( \Sim_U \right)=U$ (and consequently
  $\rank \Sim_U = \dim U$), $\Sim_U^2 = \Sim_U$, $\Sim_U + \Sim_{\dual{U}} = \Im_n$ and $\Sim_U \Sim_{\dual{U}} = \zz$ and $\Sim_U = \Sim_U^T$. 
  Finally because $\Sim_U$ is   
  symmetric, we  interpret it as an adjacency matrix 
of a weighted undirected graph. 

\begin{definition} \label{def:codegraph}
Let $U \subset \F^n$ be a linear code with a trivial hull. The \emph{weighted graph $\graph{G}_{U}$ associated to $U$}
is the graph defined by the adjacency matrix $\Sim_U$.
\end{definition}

\begin{theorem} \label{theo:red}
Let $\AC$ and $\BC$ be two linear codes  of length $n$ over a field $\F$ having both a trivial hull.
Then $\BC \sim_\pi \AC $ for some permutation $\pi$ in $\sym_n$ 
if and only if  
$\graph{G}_{\BC} \sim_\pi \graph{G}_{\AC}$.

\end{theorem}
\begin{IEEEproof}
Let assume that there exists a permutation matrix such that  $\BC = \AC \Xm$.
Since  $\BC$ has a trivial hull  the square matrix $\Sim_{\BC}$ exist and 
one has
\begin{eqnarray*}
\Sim_{\BC} & = & 
\left( \Gm_{\AC} \Xm \right)^T \Big( \Gm_{\AC} \Xm \left( \Gm_{\AC} \Xm \right)^T \Big)^{-1} 
\left( \Gm_{\AC} \Xm \right) \\
& = &  \Xm^T \Gm_{\AC} ^T \Big( \Gm_{\AC} \Xm \Xm^T \Gm_{\AC}^T \Big)^{-1} \Gm_{\AC} \Xm 
\end{eqnarray*}
Because  $\Xm\Xm^T = \Im_n$ we have 
$\Sim_{\BC}  =  \Xm^T \Sim_{\AC} \Xm$.
Consequently $\graph{G}_{\BC} $ is isomorphic to $\graph{G}_{\AC}$ by means of $\Xm$.
Reciprocally let us assume that there exists a permutation matrix $\Xm$ such that it holds
\begin{equation} \label{eq:ISOSIGMA}
\left \{
\begin{array}{lcl}
	\Sim_{\BC} &=& \Xm^T \Sim_{\AC} \Xm, \\
	\Sim_{\dual{\BC}} &=& \Xm^T \Sim_{\dual{\AC}} \Xm.
\end{array}
\right.
\end{equation} 
By definition of  $\Sim_{\AC}$ and $\Sim_{\BC}$, we know that $\im \left( \Sim_{\AC} \right) = \AC$ 
and $\im \left( \Sim_{\BC} \right) = \BC$. From \eqref{eq:ISOSIGMA} we also have  
$\im \left( \Sim_{\BC} \right) = \im \left( \Sim_{\AC} \right) \Xm$, which proves that $\BC$ and $\AC$ are permutation equivalent. 
\end{IEEEproof}
 
Note that because $\dual{\BC}$ and $\dual{\AC}$ are permutation equivalent
through the same permutation $\Xm$, we also have that $\graph{G}_{\dual{\BC}}$ and $\graph{G}_{\dual{\AC}}$ are  isomorphic.
We are now able to state the main result of the paper.

\begin{theorem} \label{theo:Genred}
There is a deterministic reduction running in $O\left ( n^\omega \right)$ time 
from deciding if two codes with trivial hulls over a field $\F$ and
length $n$ are permutation equivalent to deciding whether two weighted undirected graphs 
having $n$ vertices and weights in $\F$ are isomorphic.
\end{theorem}
 
 \begin{IEEEproof}
 The reduction comes from Theorem~\ref{theo:red}, which in turn  bowls down to computing 
 $\Sim_\AC$ and $\Sim_\BC$. This  can be done in 
 $O(n^\omega)$ operations.
 \end{IEEEproof}

 \begin{remark}
 We gathered in Table~\ref{CEP:F2} the time to solve the permutation equivalence for  binary linear codes of length $n$ and dimension $n/2$ with trivial hulls. 
 
 \begin{table}[h]
 \caption{Time in seconds to solve the permutation equivalence for randomly drawn binary codes with  a trivial hull of length $n$ and dimension $n/2$.  Results were obtained with Magma  on 
MacBookPro, Intel Core i7, 2.5 GHz, 16GB ram  
(version  2.6r7 \texttt{nauty} and \texttt{Traces} packages).} \label{CEP:F2}
    \begin{center}
        \begin{tabular}{@{}rrr@{}} 
        \toprule
          $n$ & $\Sim_\AC $  &  $\mathsf{GI}(n) $\\
	\midrule
           $1,000$     &   $0.06$   &  $0.01$ \\
           $5,000$     &    $1.6$  &   $0.11$\\
           $10,000$     &    $6.5$  &  $0.5$ \\
           $20,000$     &   $31.1$   & $1.8$  \\
           $30,000$     &    $81.1$  &  $4.1$ \\
           $40,000$     &      $153$ &  $7.5$ \\
           $50,000$     &   $283$   &   $11.9$\\
          \bottomrule
        \end{tabular}
    \end{center}
\end{table}
\end{remark}

 \section{Codes With Non-Trivial Hulls} \label{sec:non)trivial-hull}
 
 We present two approaches to extend the reduction given in Theorem~\ref{theo:Genred}
 to  codes with non trivial hulls. The first approach applies to codes defined over a field $\F = \fpm$ 
 with $m > 1$ and $p$ is a prime number as long as they have a trivial hull with respect to a Hermitian product. The second approach deals with the remaining cases, in particular  codes defined over a prime field $\fp$. In the sequel we explain in more details these two methods.
 
 \subsection{Hermitian inner product}
 In this part we consider codes that  are defined over  $\F = \fpm$ 
 with $m > 1$ and $p$ is a prime number.  The idea is to consider a Hermitian inner product  
 $ \hscp{\theta}{\xv}{\yv} \eqdef \sum_{i=1}^n x_i \theta(y_i)$ where $\theta$ is an automorphism of  $\fpm$.
 We recall that such automorphisms are of the form $z \mapsto z^{p^e}$ where $e$ lies in $\rg{0}{m-1}$.
 From now on, we consider an automorphism $\theta$. We define in the same manner the orthogonal 
 $\hdual{\theta}{U}$ of a set $U \subset \fpm^n$ as the set of vectors that are orthogonal to $U$
 with respect to $ \hscp{\theta}{\cdot}{\cdot}$. Furthermore for any permutation $\pi \in \sym_n$ and for any $\xv$, $\yv$ in $\fpm^n$ we have 
 \[
 \hscp{\theta}{\xv^\pi}{\yv^\pi}  =  \hscp{\theta}{\xv}{\yv}, 
 \]
 Hence $\AC \sim_\pi \BC$ if and only if $\hdual{\theta}{\AC} \sim_\pi \hdual{\theta}{\BC}$
 for  any linear codes $\AC \subset \fpm^n$ and $\BC \subset \fpm^n$. Following 
 \eqref{eq:SigU}  we define 
 $\Sim_U^{(\theta)}$ for a trivial-hull linear code $U \subset \fpm^n$ with respect to the hermitian inner product as
\[
\Sim_U^{(\theta)} \eqdef \theta(\Gm_U^T) \Big( \Gm_U \theta(\Gm_U^T) \Big)^{-1} \Gm_U,
\] 
with the convention that $\theta(\Zm) \eqdef \begin{bmatrix} \theta(z_{i,j}) \end{bmatrix}$ for any matrix
$\Zm = \begin{bmatrix} z_{i,j} \end{bmatrix}$. It is not difficult to see that $\Sim_U^{(\theta)}$
does not depend on $\Gm_U$ and it is possible to define a graph $\graph{G}^{(\theta)}_{U}$ with respect to $\Sim_U^{(\theta)}$ as 
in Definition~\ref{def:codegraph}.
Furthermore  since for any matrix $\Xm$ with binary entries we have 
$\theta(\Xm) = \Xm$, we are able to generalize Theorem~\ref{theo:red}. 

\begin{proposition} \label{prop:redhermitian}
Let $\AC$ and $\BC$ be two linear codes  of length $n$ over a field $\fpm$ having both a trivial hull with respect to the hermitian product $ \hscp{\theta}{\cdot}{\cdot}$ where $\theta$ is any automorphism of $\fpm$.
Then $\BC \sim_\pi \AC $ for some permutation $\pi$ in $\sym_n$ 
if and only if  
$\graph{G}^{(\theta)}_{\BC} \sim_\pi \graph{G}^{(\theta)}_{\AC}$.
\end{proposition}

 \subsection{General case}
 The result obtained in Proposition  \ref{prop:redhermitian} permits us to solve the permutation code equivalence in more cases, especially when m is very large.  The remaining  unsolved cases are therefore  codes defined over a prime field $\fp$, and those defined over $\fpm$ with $m > 1$ that have non trivial hulls with respect to any hermitian inner product over $\fpm$. We develop a general approach  based  on testing the equivalence between shortened codes. We recall that  the \emph{shortened} code of $U$ over  $\set{I} \subset \rg{1}{n}$ 
is   defined as 
 \begin{equation} \label{eq:short}
 	\sh{\set{I}}{U} \eqdef \Big \{ \uv \in U ~:~ \forall i \in \set{I}, \; \; u_i = 0 \Big \}.
\end{equation}
Moreover,  a set $\set{I} \subset \rg{1}{n}$ is an \emph{information set} for a linear code $U$ if there exists  
 a generator matrix such that its restriction to the columns that belong to $\set{I}$ is the identity matrix.  We have then the following property.

 \begin{proposition} \label{prop:shortrivial}
 If $\set{I} \subset \rg{1}{n}$ is an information set for  $U\cap \dual{U}$ then $\sh{\set{I}}{U}$ has a trivial hull.
 \end{proposition}
 \begin{IEEEproof}
 	Let us set $\hull{U} \eqdef U\cap \dual{U}$.
 	Firstly, we have necessarily $U =  \sh{\set{I}}{U} \oplus \hull{U}$
	 by considering for instance a generator matrix of $U$ in systematic form over $\set{I}$ 
	 and such that the first rows form a 	basis of  $\hull{U}$. The last rows form therefore a basis 
	of  $\sh{\set{I}}{U}$ and consequently $\sh{\set{I}}{U} \cap \hull{U} = \{\zz \}$.
	
	Let us assume now that $\vv$ belongs to $\sh{\set{I}}{U} \cap \dualp{\sh{\set{I}}{U}}$.
	        Then
	       $\vv$  is also orthogonal to $\hull{U}$ since
	       $\vv \in \sh{\set{I}}{U} \subset U$, 
	       and therefore $\vv$ is orthogonal
        to $\sh{\set{I}}{U} + \hull{U}  = U$. This implies that $\vv$ belongs to $U \cap \dual{U} = \hull{U}$.
        This is possible if and only if $\vv = \zz$ which proves that the hull of $\sh{\set{I}}{U}$ is trivial.	
 \end{IEEEproof}

\begin{theorem} \label{theo:nontrivialhull}
The permutation code equivalence between two linear codes of length $n$ over a field $\F$
can  be decided in
 $O\Big(   h n^{\omega+h+1} \GI(n)     \Big)$ 
operations where $h$ is the dimension of the hull, and
$\GI(n)$  is  the time complexity for deciding the isomorphism of weighted graphs with $n$ vertices and weights in $\F$.
\end{theorem}
 
\begin{IEEEproof}
We assume that we have at our disposal a deterministic algorithm   that solves the  weighted graph isomorphism. Thanks to the reduction given in Theorem~\ref{theo:Genred},  we have therefore
a deterministic algorithm  that  tests if two codes with trivial hulls are equivalent or not.

It is not difficult to see that  for any linear code $U \subset \F^n$  
and for any set $\set{\set{I}}$ as  in
Proposition~\ref{prop:shortrivial} it holds then  
\begin{equation} \label{eq:shortequiv}
\forall \pi \in \sym_n, \;\; 
\sh{\set{I}^\pi}{U^\pi} \sim_\pi  \sh{\set{I}}{U}.
\end{equation}
This suggests searching for $\set{I}^\pi$ among the sets $\set{J} \subset \rg{1}{n}$ of cardinality $h$ 
 such that $\sh{\set{J}}{U^\pi}$ has a trivial hull.
However it may happen that two inequivalent codes have equivalent shortened codes.
A way to deal with this issue is to observe that  for  all
 $\ell \in\rg{1}{n} \setminus \set{I}$  
  the codes $\sh{\{ \ell\} \cup \set{I} \setminus \{ i \} }{U}$   and
$\sh{ \{ \pi(\ell) \} \cup \set{I}^\pi \setminus \{ \pi(i) \}}{U^\pi}$ are equivalent
  for each $i \in \set{I}$.
  Note that the recourse to an integer $\ell \notin \set{I}$  is necessary because our test requires to have codes shortened on sets of size $\dim   U \cap \dual{U} $.

 We  now describe a method for testing if 
  two linear codes $\AC$ and $\BC$ of length $n$ both having hulls of dimension $h$ are equivalent or not.
Firstly we fix a set  $\set{I} \subset \rg{1}{n}$  
and an integer $\ell \in \rg{1}{n} \setminus \set{I}$
such that $\sh{\set{I}}{\AC}$ has a trivial hull. 
Next, we search  for a set $\set{J} \subset \rg{1}{n}$ of cardinality of $h$
such that $\sh{\set{J}}{\BC}$ has a trivial hull, an integer $\ell' \in \rg{1}{n} \setminus \set{J}$
 and a permutation 
$\gamma : \set{I}  \rightarrow  \set{J}  $ such that the following holds
 \begin{enumerate}
 	\item  $\sh{\set{I}}{\AC}$ and $\sh{\set{J}}{\BC}$ are equivalent which can be decided in $O\Big( (n-h)^w \GI\left(n-h \right)\Big)$ operations.
	\item   $\sh{\set{I}}{\AC}$ punctured at $\ell$ is equivalent to  $\sh{\set{J}}{\BC}$ punctured at $\ell'$
	 This step  can be done with at most $O\Big( (n-1-h)^w \GI\left(n-1-h \right)\Big)$ operations.
	\item For $i \in \set{I}$, $\sh{\{ \ell\} \cup \set{I} \setminus \{ i \} }{\AC}$   and
$\sh{ \{ \ell' \} \cup \set{J} \setminus \{ \gamma(i) \}}{\BC}$ are equivalent. 
This step requires  $O\Big( (n-h)^w \GI\left(n-h \right)\Big)$ operations for each $i$.
 \end{enumerate}
 When $\AC$ and $\BC$ are equivalent, this procedure will necessarily find the quantities 
 $\set{J}$, $\ell'$ and $\gamma$,  unlike the case where $\AC$ and $\BC$ are inequivalent codes.
 The (worst-case) cost of this approach is therefore  
 $O\Big(   \binom{n}{h} h! (n-h)^{\omega+1} (h + 2)\GI(n-h)     \Big)$ 
 operations, which asymptotically  gives $O\Big(   h n^{\omega+h+1} \GI(n)     \Big)$.
 \end{IEEEproof}

\section*{Acknowledgment}

This work has been partially supported by  MANTA (ANR-15-CE39-0013), 
CBCRYPT (ANR-17-CE39-0007) and  the MOUSTIC project with the support from the European
Regional Development Fund (ERDF) and  the Regional Council of Normandie.

\bibliographystyle{alpha}
\bibliography{codecrypto,graph}

\end{document}